\RequirePackage{fix-cm}
\documentclass[twocolumn]{svjour3}          
\smartqed  
\usepackage{graphicx, color}
\journalname{J Sol-Gel Sci Technol}
\begin{document}

\title{Ultralow-density double-layer silica aerogel fabrication for the intact capture of cosmic dust in low-Earth orbits
}

\titlerunning{Ultralow-density double-layer silica aerogel fabrication}        

\author{Makoto Tabata \and
	Hideyuki Kawai \and
	Hajime Yano \and
	Eiichi Imai \and
	Hirofumi Hashimoto \and
	Shin-ichi Yokobori \and
       Akihiko Yamagishi 
}


\institute{M. Tabata \at
              Department of Physics, Chiba University, Chiba 263-8522, Japan\\Institute of Space and Astronautical Science (ISAS), Japan Aerospace Exploration Agency (JAXA), Sagamihara 252-5210, Japan\\
              \email{makoto@hepburn.s.chiba-u.ac.jp}           
           \and
           H. Kawai \at
              Department of Physics, Chiba University, Chiba 263-8522, Japan
	    \and
           H. Yano \and H. Hashimoto \at
              Institute of Space and Astronautical Science (ISAS), Japan Aerospace Exploration Agency (JAXA), Sagamihara 252-5210, Japan
	    \and
           E. Imai \at
              Department of Bioengineering, Nagaoka University of Technology, Nagaoka 940-2188, Japan
	    \and
           S. Yokobori \and A. Yamagishi \at
              Department of Applied Life Sciences, Tokyo University of Pharmacy and Life Sciences, Hachioji 192-0392, Japan
}

\date{Received: date / Accepted: date}

\maketitle

\begin{abstract}
The fabrication of an ultralow-density hydrophobic silica aerogel for the intact capture cosmic dust during the Tanpopo mission is described. The Tanpopo experiment performed on the International Space Station orbiting the Earth includes the collection of terrestrial and interplanetary dust samples on a silica aerogel capture medium exposed to space for later ground-based biological and chemical analyses. The key to the mission's success is the development of high-performance capture media, and the major challenge is to satisfy the mechanical requirements as a spacecraft payload while maximizing the performance for intact capture. To this end, an ultralow-density (0.01 g cm$^{-3}$) soft aerogel was employed in combination with a relatively robust 0.03 g cm$^{-3}$ aerogel. A procedure was also established for the mass production of double-layer aerogel tiles formed with a 0.01 g cm$^{-3}$ surface layer and a 0.03 g cm$^{-3}$ open-topped, box-shaped base layer, and 60 aerogel tiles were manufactured. The fabricated aerogel tiles have been demonstrated to be suitable as flight hardware with respect to both scientific and safety requirements.
\end{abstract}

\section{Introduction}
\label{sec1}

Silica aerogels \cite{Ref1} are well known as highly porous substances comprising silicon dioxide clusters and air-filled open pores in the order of tens of nanometers. They are also distinguished by their optical transparency. The production of silica aerogels involves a two-step process. A wet silica gel is first synthesized via sol--gel polymerization, and then the dry gel is obtained via supercritical drying using, for example, alcohol or carbon dioxide as the supercritical fluid or via ambient pressure drying (e.g., \cite{Ref2}). One of the most striking characteristics of silica aerogels is their density ($\rho $), which can be tuned during the production process by varying the ratio of the silica precursor to the solvent in the wet-gel synthesis step. Precise control of the density can be achieved by taking into account the shrinkage of the aerogel volume during the production process. Original silica aerogels were hydrophilic; however, it has been shown that hydrophobic aerogels can be obtained via modification of hydroxyl groups on the surface of the silica clusters \cite{Ref3}. Recently, a method for the production of hydrophobic silica aerogels was reported that provides control of the density over a wide range from 0.01 to 1.0 g cm$^{-3}$ \cite{Ref4}.

Low-density silica aerogels have been widely used in various space experiments as cosmic dust capture media \cite{Ref5}, both in low-Earth orbits (LEOs) \cite{Ref6,Ref7} and deep space \cite{Ref8}. The capture of cosmic dust in space involves impact between the capture medium and dust grains with a size of tens of micrometers at a hypervelocity of several kilometers per second. Therefore, extremely low-density media (e.g., less than 0.1 g cm$^{-3}$) are required to capture intact dust grains with no evaporation due to impact shock. Silica aerogels are one of the most promising types of capture media due to their low-density porous characteristics \cite{Ref6}. Dust grains collide with the aerogel surface, partially penetrate the aerogel thickness to form an impact cavity (also called an \textit{impact track}), and eventually come to rest. The optical transparency of silica aerogels is also advantageous because the impact tracks can be visibly detected thereby allowing identification of the dust grains trapped inside with the aid of microscopes.

Cosmic dust samples returned to laboratories on the ground for analysis using state-of-the-art techniques are of importance in astrobiology, planetary science, and space debris research. In 2007, the Tanpopo project was proposed as Japan's astrobiological space mission comprising composite experiments to be performed aboard the International Space Station (ISS) in LEOs \cite{Ref9,Ref10}. One of the experiments involves exposure of silica aerogel to space for the collection of cosmic dust, including natural terrestrial dust grains, interplanetary dust particles, possible interstellar dust particles, and artificial space debris. The Tanpopo mission will explore the existence of terrestrial microbes in LEOs and examine the prebiotic chemistry of natural dust provided from interplanetary space to the Earth. Moreover, the experiment will provide information on the space debris environment in LEOs. To achieve these objectives, hydrophobic silica aerogels with $\rho $ = 0.01 g cm$^{-3}$ were developed for this project. Herein, we describe the results of our study on the fabrication of a flight model silica aerogel (hereafter referred to simply as \textit{aerogel}).

\section{Requirements for the Tanpopo aerogel}
\label{sec2}

For successful analysis of the cosmic dust grains captured during the Tanpopo experiment using a low-density aerogel, it is crucial to reduce the degradation of the particles due to impact shock and heat during momentum reduction. Previously we succeeded in reliably producing hydrophobic and monolithic aerogel tiles with $\rho $ = 0.01 g cm$^{-3}$ \cite{Ref4}. We planned to use this 0.01 g cm$^{-3}$ aerogel at least as the upper-most layer of the aerogel capture medium \cite{Ref9,Ref10,Ref11}.

The dimensions for one Tanpopo capture panel module are 100 $\times $ 100 $\times $ 20 mm$^3$. Each capture panel comprises an aerogel tile and its holder made of an aluminum alloy. Thus, the inner width and thickness for each aerogel tile are 94 mm and 17 mm, respectively. The panels will be mounted on the Exposed Experiment Handrail Attachment Mechanism (ExHAM), newly developed by the Japan Aerospace Exploration Agency (JAXA), and attached to the Exposed Facility in the Japanese Experiment Module (JEM) on the ISS using a robotic arm and airlock. Four aerogel tiles will be attached to each of three different sides (i.e., the East ($X$), which is the direction of the ISS orbit, North ($Y$), and zenith ($Z$) directions) for approximately one year. The one-year exposure of the aerogel tiles will be repeated three times by replacing the aerogel tiles with new ones. Therefore, a total of 36 aerogel tiles (i.e., 4 tiles $\times $ 3 directions $\times $ 3 years) and spare tiles for use in ground-based tests must be fabricated.

The aerogel tiles should resist rocket launch vibrations, because the capture panels will be transported to the ISS by a supply spacecraft. To protect the extremely fragile 0.01 g cm$^{-3}$ aerogel from launch vibrations, a prototype box-framing aerogel was designed \cite{Ref12} that consisted of two layers; i.e., upper (surface) and lower (base) layers with $\rho $ = 0.01 and 0.03 g cm$^{-3}$, respectively. The 0.03 g cm$^{-3}$ base layer was formed as an open-topped box shape, and the 0.01 g cm$^{-3}$ surface layer was integrated into the box. The prototype box-framing aerogel was installed in the holder by fixing the 0.03 g cm$^{-3}$ frame between the holder body and the lid. No serious damage was reported in vibration tests, because the 0.01 g cm$^{-3}$ surface layer never touched the holder \cite{Ref12}. We thus chose to fabricate the box-framing aerogel tile in the present study.

\section{Methodology}
\label{sec:3}

\subsection{Raw materials}
\label{sec:3.1}

Table \ref{tab:1} lists the raw materials and quantities used for producing three box-framing aerogel tiles. The sols were prepared according to the formulations in Table \ref{tab:1}, and then the appropriate volumes or weights were distributed to each of the three molds. Any remaining sol was discarded.

\begin{table*}
\caption{Chemicals used for the synthesis of three box-framing wet gels}
\label{tab:1}       
\begin{tabular}{lll}
\hline\noalign{\smallskip}
Chemicals & Dose [g] for base (surface) layer & Supplier \\
\noalign{\smallskip}\hline\noalign{\smallskip}
Polymethoxy siloxane & 11.84 (2.28) & Fuso Chemical Co., Ltd., Japan \\
Ethanol (99.5) for HPLC$^a$ & 218.86 (149.02) & Kanto Chemical Co., Inc., Japan \\
28\% Ammonia solution & 11.27 (8.24) & Wako Pure Chemical Industries, Ltd., Japan \\
\noalign{\smallskip}\hline
$^a$ High-performance liquid chromatography
\end{tabular}
\end{table*}

\subsection{Fabrication procedure}
\label{sec:3.2}

\begin{figure}
  \includegraphics[width=0.48\textwidth]{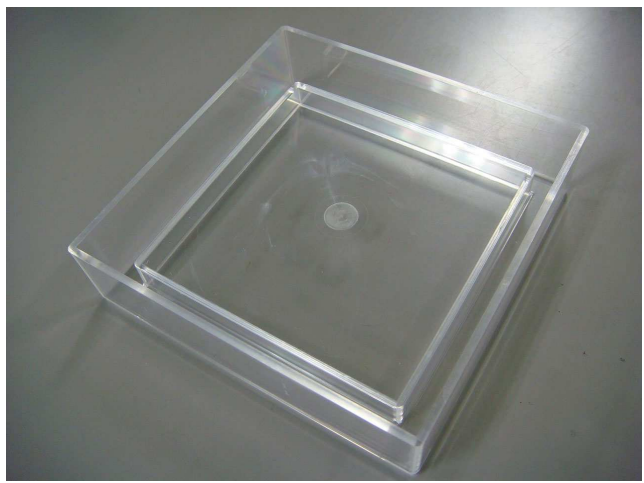}
\caption{Custom-made polystyrene mold for the production of a box-framing aerogel tile. The inner width is 96 mm and the inner depth is 25 mm. The mold also contains a convex tableland with dimensions of 82 $\times $ 82 $\times $ 10.5 mm$^3$}
\label{fig:1}
\end{figure}

A custom-made special mold fabricated from polystyrene (PS) was used (Fig. \ref{fig:1}) to form the box-framing wet gel structure. This mold was used for modeling the upside-down base-layer wet gel with $\rho $ = 0.03 g cm$^{-3}$. It was basically an open-topped square box shape with an inner width of 96 mm. A key feature of the mold was a square-shaped convex area on the inner bottom plate with a width of 82 mm and a height of 10.5 mm, which provided a space with dimensions of 82 $\times $ 82 $\times $ 10.5 mm$^3$ for pouring the sol prepared for the 0.01 g cm$^{-3}$ surface layer wet gel into the 0.03 g cm$^{-3}$ base layer after the base wet gel block was detached from the mold by tuning it.

The wet gel was produced via a successive sol--gel condensation polymerization reaction:
\[ \rm{CH_3O[Si(OCH_3)_2O]_\textit{n}CH_3 + (\textit{n} + 1)H_2O} \]
\[ \to \rm{(SiO_2)_\textit{n} + (2\textit{n} + 2)CH_3OH.} \]
First, wet gel blocks for the 0.03 g cm$^{-3}$ base layers were synthesized. In a polypropylene or polyethylene beaker, solution A was prepared by mixing polymethoxy siloxane with high-performance liquid chromatography (HPLC)-grade ethanol using half the amount listed in Table \ref{tab:1}. Solution B was then prepared by mixing aqueous ammonia with the remaining amount of ethanol in a second beaker. Solution B was then poured into solution A, and the resulting mixture was strongly stirred for 30 s. The mixture was then sequentially poured into the three molds, with measurement of the solution level using a ruler. The target height of the mixed solution (i.e., the thickness of each wet gel block) was 18 mm. At room temperature (20$^\circ $C--21$^\circ $C), gelation was complete after approximately three minutes (3:00--3:30) from the time solution B was poured into solution A. After a further five minutes, the surfaces of the wet gel blocks in the molds were covered with HPLC-grade ethanol (4 mL) in order to prevent them from drying. The molds were then covered with PS lids from commercially available box-shaped cases (described below) and aged in a sealed, stainless-steel tank for 1--2 days at room temperature.

\begin{figure}
  \includegraphics[width=0.48\textwidth]{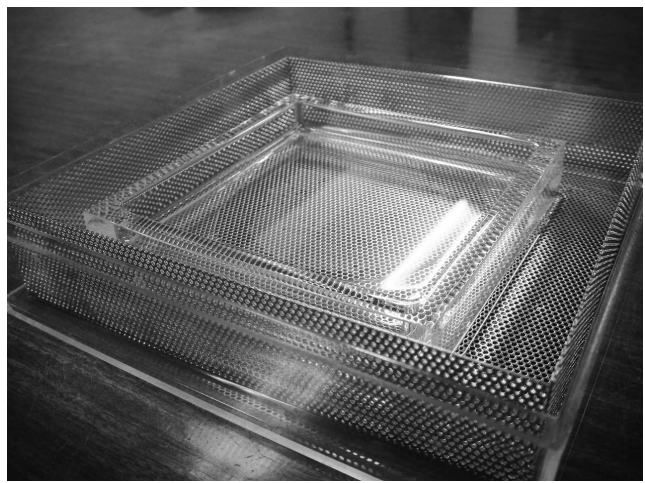}
\caption{Base layer wet gel for the 0.03 g cm$^{-3}$ aerogel layer detached from a custom-made PS mold into a punched tray placed in a PS case. This wet gel sample was experimentally manufactured in an uncontrolled environment designed to prevent contamination}
\label{fig:2}
\end{figure}

Next, wet gels for the 0.01 g cm$^{-3}$ surface layers were synthesized on the base layers. Prior to detachment of the wet gel blocks for the 0.03 g cm$^{-3}$ base layers from the PS molds, stainless steel punched trays with the dimensions of 112 $\times $ 112 $\times $ 20 mm$^3$ were placed in square PS cases with outer dimensions of 120 $\times $ 120 $\times $ 32 mm$^3$, including the lids used in the previous process. To prevent drying of the wet gels, HPLC-grade ethanol (4 mL) was dropped in each PS case in advance in order to fill the cases with ethanol vapor. As shown in Fig. \ref{fig:2}, each base layer wet gel block was detached from the mold by overturning it on the punched tray and then temporarily covered with the PS lid. The sol for the 0.01 g cm$^{-3}$ surface layers was prepared using the same procedure as described above for the base layers, and after briefly removing the lid from each case, 52.0 $\pm $ 0.2 g of the sol was poured into the hollow created in each base layer by the custom-made PS mold; i.e., the wet gel block for the 0.03 g cm$^{-3}$ base layer functioned as a mold for the surface layer wet gel. Each case was again covered with a lid to prevent the sol from drying. Three wet gel blocks were processed at the same time. The surface layer sols gelled at room temperature (20$^\circ $C--22$^\circ $C) within approximately 15 min from the time solution B was mixed with solution A, and then the cases were transferred to the sealed tank.

Wet gel aging was conducted as follows. Each wet gel was first statically aged in the sealed tank at room temperature for approximately five days. Then, to further strengthen the silica networks, second-stage aging was performed for approximately six days at 35$^\circ $C using a water bath. In this aging stage, each wet gel on a punched tray was separated from the PS case and fully sunk into HPLC-grade ethanol to prevent drying. The sealed tank filled with ethanol and containing the wet gel was then immersed into a water bath.

After retrieving the tank from the water bath, the wet gels were subjected to special treatment and post-processing in order to render them hydrophobic. In the hydrophobic treatment reaction, hydroxyl groups ($-$OH) on the surfaces of the silica clusters were replaced with trimethylsiloxy groups [$-$OSi(CH$_3$)$_3$]. The wet gels on the punched trays were temporarily extracted from the tank filled with ethanol. Hexamethyldisilazane\newline [((CH$_3$)$_3$Si)$_2$NH; Dow Corning$^{\textregistered }$ Z-6079 Silazane; Dow Corning Toray Co., Ltd., Japan] (500 mL) was then poured into the ethanol (approximately 3,300 mL) in the tank, and the solution was stirred. Six wet gel tiles were statically soaked in the silazane solution at room temperature for approximately three days. Then, to reduce the impurities in the silazane solution, including the ammonia generated in the hydrophobic treatment, residual silazane, and residual products in the sol--gel polymerization, the silazane solution was replaced with new ethanol three times at intervals of more than one day.

Finally, the wet gel was subjected to supercritical carbon dioxide (CO$_2$) drying using a homebuilt supercritical CO$_2$ extraction system. The temperature and pressure at the critical point of CO$_2$ are 31$^\circ $C and 7.4 MPa, respectively \cite{Ref13}. Although it would be ideal to replace the ethanol in the wet gel with liquefied CO$_2$ prior to CO$_2$ extraction at the supercritical phase, it was more practical to replace the solvent and perform the extraction simultaneously. The system was equipped with a 7.6 L pressure vessel (autoclave; Nitto Koatsu Co., Ltd., Japan) with a depth of 30 cm, which was filled with new HPLC-grade ethanol. A total of 12 wet gel tiles were stored in the autoclave by directly stacking the punched trays in the ethanol, and then autoclave was sealed.

The process was begun at room temperature by slightly opening the CO$_2$ input valve to gradually supply CO$_2$ gas from a liquefied CO$_2$ cylinder (Showa Denko Gas Products Co., Ltd., Japan) to the autoclave. This input valve remained open until the mixing of ethanol and liquefied CO$_2$ reached equilibrium at a pressure of approximately 5.5 MPa. The tiles were left under these conditions for at least 12 h, and then the pressure in the autoclave was raised to 9 MPa by injecting liquefied CO$_2$, typically over 10 min, using a pump [NP-D-701(J); Nihon Seimitsu Kagaku Co., Ltd., Japan]. Once the pump was halted, the temperature in the autoclave was raised to 40$^\circ $C over approximately 2 h using a heater surrounding the autoclave. During heating, the pressure was maintained below 11 MPa (i.e., the safety pressure of the autoclave), by extracting the ethanol/CO$_2$ mixture through an output valve on the autoclave.

Liquefied CO$_2$ was then injected for 19 h by continually operating the pump at a rate of 29 mL/min, followed by heating to 80$^\circ $C over 8 h with further liquefied CO$_2$ injection. The pressure was again maintained below 11 MPa by continually extracting the solvent. After heating under these CO$_2$-rich conditions for approximately 2 h, the completion of solvent extraction was confirmed by determining the ethanol concentration in the extracted gas, which was collected in a plastic bag. The ethanol concentration was measured using a gas sampling pump kit with an ethanol inspection tube (Gastec Corporation, Japan). When the obtained value was $<$200 ppm, the pump was halted and the input valve was closed. Next, while maintaining a high temperature above the critical temperature, the pressure was slowly reduced to atmospheric pressure over approximately 14 h. Finally, the autoclave was opened at room temperature, and the dry gel tile was retrieved. The gas-phase CO$_2$ in the dry gel was spontaneously substituted by atmospheric air, providing the final box-framing aerogel product.

\section{Results}
\label{sec:4}

\subsection{Mass production}
\label{sec:4.1}

\begin{figure}
  \includegraphics[width=0.48\textwidth]{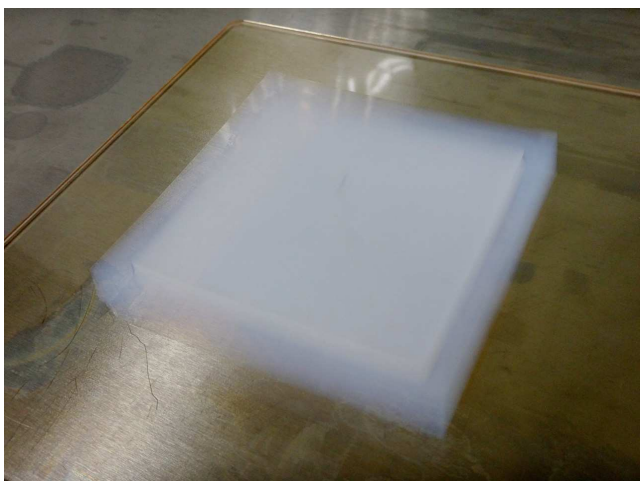}
\caption{Box-framing aerogel tile with approximate dimensions of 87 $\times $ 87 $\times $ 17 mm$^3$}
\label{fig:3}
\end{figure}

A total of 60 box-framing aerogel tiles were fabricated in five production batches (12 tiles per batch) at Chiba University as actual capture panel components in a controlled environment designed to prevent contamination. To effectively employ all resources, a new wet gel synthesis was initiated on a weekly basis. After five operations of the supercritical drying apparatus, all of the 60 dry gel tiles were obtained. Only three tiles were slightly damaged; thus, a total of 57 flight-qualified aerogel tiles were fabricated (see Fig. \ref{fig:3}).

\subsection{Determination of the tile dimensions}
\label{sec:4.2}

\begin{figure}
  \includegraphics[width=0.48\textwidth]{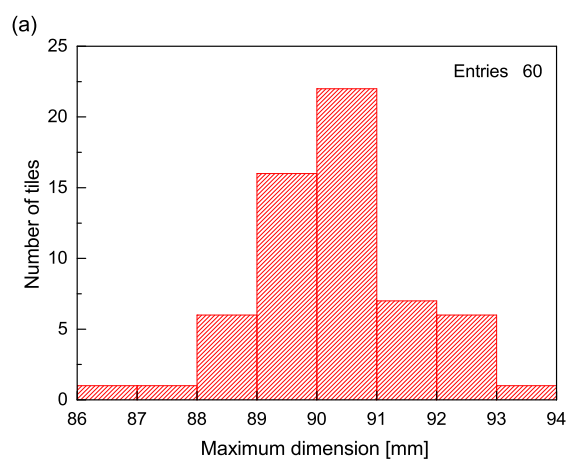}
  \includegraphics[width=0.48\textwidth]{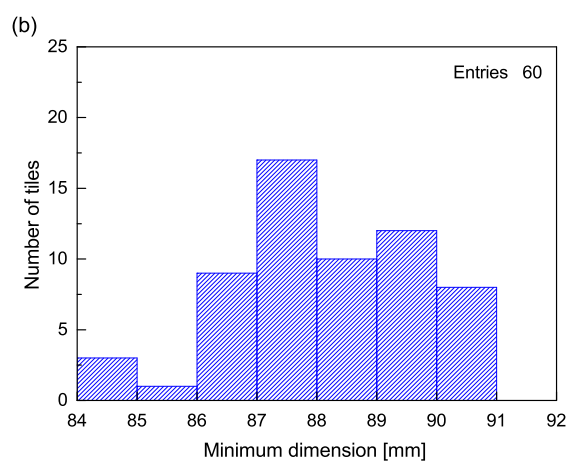}
\caption{Distribution of the (a) maximum and (b) minimum widths of the side walls for the 60 aerogel tiles}
\label{fig:4}
\end{figure}

\begin{figure}
  \includegraphics[width=0.48\textwidth]{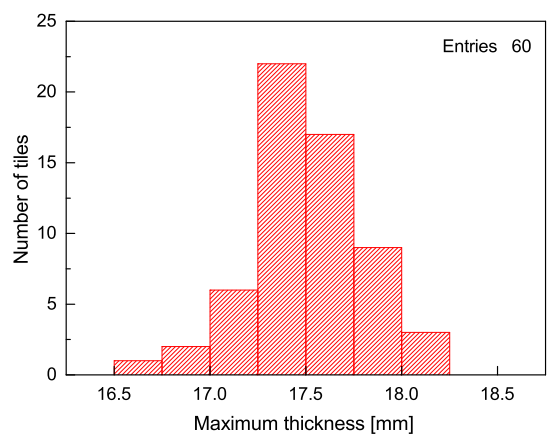}
\caption{Distribution of the maximum thickness at the tile corners for the 60 aerogel tiles}
\label{fig:5}
\end{figure}

The characteristic dimensions (i.e., the maximum and minimum length/thickness of each side wall) of all of the fabricated aerogel tiles were measured using a ruler. The width of the custom-made mold slightly tapered toward the bottom to facilitate the manufacture of the mold itself and the detachment of the formed wet gel. Thus, the wet gel block for the 0.03 g cm$^{-3}$ base layer detached upside down from the mold had maximum and minimum widths at the bottom and top sides of each side wall, respectively. The same was true on the dry gel block. Fig. \ref{fig:4} shows the distribution of the maximum and minimum widths for the 60 aerogel tiles. Likewise, Fig. \ref{fig:5} presents the distribution of the maximum thickness. The maximum thickness was obtained at the tile corners due to the meniscus geometry of the surfaces of the wet gel blocks for the base layers. For the 0.01 g cm$^{-3}$ surface layers, the surface levels were below the levels of the 0.03 g cm$^{-3}$ base layers for all of the aerogel tiles. This design was used to protect the inner layer from rocket launch vibrations.

\subsection{Indirect density analysis}
\label{sec:4.3}

The bulk density of each aerogel tile was calculated by dividing its mass by its volume. However, it was difficult to separately determine the values for each layer of the box-framing aerogel tiles because they were formed as chemically combined, monolithic tiles. Therefore, instead of direct measurement of the aerogel densities, the refractive index of each box-framing aerogel tile was determined and then converted to the density based on Eq. (\ref{eq:1}) in the Discussion section. The prism minimum deviation method (also referred to as the Fraunhofer method \cite{Ref14}.) was used to determine the refractive indices of the aerogel tiles using a laser. The refractive index at the corner of one of the mass-produced box-framing aerogel tiles with a typical size was determined. The minimum deviation of the laser beam was measured on a screen approximately 1.8 m downstream of the aerogel tile. First, the refractive index of the base layer was found to be 1.00986 using two lasers with wavelengths ($\lambda $) of 405 and 633 nm (i.e., the identical value was obtained using both the lasers). A corner of the surface layer was then exposed by partially removing the base layer with a knife at a point where the surface layer had been partially separated from the base layer due to extra shrinkage of the surface layer. The refractive index of the surface layer was found to be 1.0030 using the laser with $\lambda $ = 633 nm (the laser beam with $\lambda $ = 405 nm did not penetrate due to the low transmittance of the surface layer). For the base and surface layers, the measurement errors derived from the laser spot size observed on the screen were $\pm $0.0008 and $\pm $0.0011 at $\lambda $ = 633 nm, respectively. The spot size depended on the degree of light scattering in the aerogel tile.

\section{Discussion}
\label{sec:5}

\subsection{Fabrication technology}
\label{sec:5.1}

A novel method (hereafter referred to as the classic KEK method) independently developed at the High Energy Accelerator Research Organization (KEK) in Japan for the production of transparent, low-density, and hydrophobic aerogels was reported in Ref. \cite{Ref15}. Using this method, aerogel tiles with a refractive index of 1.013 (i.e., a density of approximately 0.048 g cm$^{-3}$) were prepared. The empirical law for conversion from the density to the refractive index ($n$) for these hydrophobic aerogel tiles is described as follows:
\begin{equation}
\label{eq:1}
n = 1.001 + 0.251\rho,
\end{equation}
where the refractive index was measured using a laser with a wavelength of 405 nm \cite{Ref14}. At that time, aerogel tiles with a refractive index of 1.010 (density of approximately 0.036 g cm$^{-3}$) could also be produced using the classic KEK method \cite{Ref16}. Modern techniques for producing more transparent, high-density aerogels have also been developed \cite{Ref4,Ref17} and standardized \cite{Ref14,Ref18}; however, the classic KEK method remains an effective technique for the production of ultralow-density aerogels \cite{Ref4}.

In the classic KEK method and related techniques, polymethoxy siloxane (methyl silicate 51; Fuso Chemical Co., Ltd., Japan) is used as the precursor for the sol--gel polymerization to simplify the wet-gel synthesis procedure. Methyl silicate 51 is prepared by condensing tetramethyl orthosilicate (TMOS), which results in oligomerization to the tetramer (on average) and results in a high silica content of 51 wt\%. Methyl silicate 51 can be readily hydrolyzed and condensed in water to produce silica sols. In the present study, aqueous ammonia was used as the water source. The use of methyl silicate precursor rather than TMOS provides aerogels with physical advantages \cite{Ref19} via a synthetic procedure that is as simple as the single-step method. In addition, this method simplifies the spontaneous formation of chemical bonds at the boundary between the 0.01 g cm$^{-3}$ surface and 0.03 g cm$^{-3}$ base layers in the wet-gel synthesis stage to generate a monolithic aerogel tile \cite{Ref20}.

The mold detachment procedure and wet gel handling are crucial for the successful preparation of ultralow-density aerogels because they can be easily damaged by their own weight or pressure from liquid solvents. Therefore, the wet gels were formed using a mold made of PS so that after aging they could be easily detached from the mold for later processing (i.e., hydrophobic treatment and supercritical drying). They were also carefully handled in and out of the ethanol solutions.

\subsection{Low density}
\label{sec:5.2}

The volume ratio of silica skeleton to pore air determines the aerogel density. This silica volume ratio can be tuned by varying the mixing ratio of the raw chemicals for the wet gel synthesis. Therefore, disregarding any volume shrinkage, the density of an aerogel can be roughly estimated in advance based on the mixing ratio. The formulation also impacts the optical transparence of aerogels. As part of our research on aerogels, we have empirically examined the mixing recipe to both adjust the aerogel density and improve the transparency of aerogels. 

With respect to the fabricated box-framing aerogel tiles, the measured density of the surface and base layers for a representative tile were 0.008 and 0.035 g cm$^{-3}$, respectively, using Eq. (\ref{eq:1}) and the measured refractive index values. The systematic measurement errors derived from the laser spot size were estimated to be $\pm $0.004 and $\pm $0.003 g cm$^{-3}$ for the surface and base layers, respectively. Although the conversion law was empirically obtained by analyzing the data for single-layer aerogel tiles with densities ranging from 0.02 to 0.06 g cm$^{-3}$ \cite{Ref14}, it was extrapolated to the lower density range. The conversion law is based on the Lorentz--Lorenz formula \cite{Ref21}, and the proportional relationship can be empirically applied over a wide range of densities \cite{Ref14}. Consequently, a nominal density of 0.01 g cm$^{-3}$ was obtained for the surface layer, whereas a slight density shift from the nominal density (i.e., 0.03 g cm$^{-3}$) was observed for the base layer; however, this shift will not degrade the performance of the aerogel tiles with respect to the capture of intact dust particles.

Lower density aerogels function well for the capture of nearly intact hypervelocity dust \cite{Ref6}. However, to further improve the performance, it is essential to decrease the pressure applied to the dust particles during the initial impact stage by reducing the density of the aerogel surface. Assuming that the density of the silica clusters in an aerogel is 2.7 g cm$^{-3}$ (identical to quartz), the porosity of the aerogel with $\rho $ = 0.01 g cm$^{-3}$ is 99.7\%. This aerogel density is one of the lowest densities used in space experiments to date.

In addition, the hydrophobic characteristic of the aerogel will ensure that the low density is maintained during the long experimental period. Many previous non-Japanese dust capture missions have used hydrophilic aerogels, which can result in apparent density changes due to age-related moisture adsorption in the ground environment and desorption in the space vacuum. These density changes can affect the intact capture performance and morphological analysis of the impact tracks. In contrast, the hydrophobic aerogel used in the Tanpopo experiment will not actively adsorb the environmental moisture or other ions. This moisture and ions can also contain contaminants, which are important to avoid because the aerogel is handled by astronauts in the pressurized module of the ISS.

\subsection{Limited tile dimensions}
\label{sec:5.3}

For the Tanpopo experiment, the permitted thickness of the aerogel tile including its holder was 2 cm. This design requirement originated from the space available on the ExHAM for storage of the capture panels. It was necessary to design the holder and density configuration for the aerogel tiles and the capture panel to these specifications while maximizing the capture performance. The length of the impact tracks created inside aerogels by cosmic dust increases as the aerogel density decreases \cite{Ref22}. To ensure capture of hypervelocity dust grains inside the aerogel, one solution is to simply increase the aerogel thickness. The use of a dense aerogel is another solution, but will result in a drop in the intact capture performance. A third possibility is to use gradient density aerogel. This approach was used on the National Aeronautics and Space Administration (NASA) Stardust spacecraft, which retrieved cometary dust samples from the comet Wild 2 \cite{Ref8}. In LEOs, the Tanpopo experiment targets cosmic dust with a diameter of several tens of micrometers and an orbiting velocity relative to the ISS (approximately 16 km/s at maximum) from all possible dust sources. In contrast, the Stardust aerogel captured cometary dust from a known object at an encounter speed of 6.1 km/s; however, the dust size was not precisely known in advance. A gradient density aerogel block was used in which the density increased from 0.01 g cm$^{-3}$ at the aerogel surface to 0.05 g cm$^{-3}$ at the bottom \cite{Ref23}.

However, it is easier to produce a simple two-layer aerogel tile than to synthesize a gradient density aerogel. In addition, tiles consisting of layers with different densities that are clearly separated by a boundary plane should increase the ability to analyze the morphologies of impact tracks, which provide physical parameters on the impact phenomena. Therefore, an aerogel tile made from two layers with different densities was developed for the Tanpopo experiment \cite{Ref12}. The nominal densities of the upper and lower layers of the aerogel tile were 0.01 and 0.03 g cm$^{-3}$, respectively. A density of 0.01 g cm$^{-3}$ for the surface layer is was near the lowest value possible with the classic KEK method. Conversely, the performance of aerogels with a density of 0.03 g cm$^{-3}$ has been proven in space; they were used in the previous Micro-Particles Capturer experiment conducted by JAXA \cite{Ref7}.

Importantly, all the 60 mass-produced aerogel tiles met the dimensional requirements for the capture panel holder. The secured area for the aerogel tile was 94 $\times $ 94 mm$^2$, and the target aerogel width was designed to be 92 mm for smooth installation into the holder. The maximum width of the aerogel tiles varied from 86.75 to 93.5 mm (Fig. \ref{fig:4}a), and therefore all of the tiles could be inserted into the holder. Note that each aerogel tile had to be fixed between the holder body and its lid. An area for compression of the edges of the base layer (0.03 g cm$^{-3}$) was created inside the lid on all four sides. For correct operation of the aerogel fixture mechanism, the acceptable maximum aerogel thickness at the tile corners is 16.5--18.0 mm, and the minimum tile width must be greater than 84 mm. The measured maximum thicknesses and minimum widths ranged from 16.5 to 18 mm (Fig. \ref{fig:5}) and 84.25 to 90.5 mm (Fig. \ref{fig:4}b), respectively.

\subsection{Robustness}
\label{sec:5.4}

Pure silica aerogels are essentially brittle media, and thus sufficient robustness is lacking, particularly for the aerogels with a density of 0.01 g cm$^{-3}$. Therefore, the 0.01 g cm$^{-3}$ aerogels must be handled very carefully. This issue was somewhat addressed by combining the very low-density aerogel with the 0.03 g cm$^{-3}$ aerogel. In 2012, a random vibration preliminary test simulating a rocket launch was performed using an experimentally-produced simple (not box framing), two-layer monolithic aerogel tile and its holder at the Space Plasma Laboratory of the Institute of Space and Astronautical Science (ISAS), JAXA. The aerogel holder consists of a body case and open-topped lid. The lid has grids to prevent the aerogel tile from escaping the holder if the tile disintegrates and to prevent the ISS crews from touching the aerogel surface. The aerogel was seriously damaged during this vibration test because it touching the grids \cite{Ref12}.

To resolve this problem, the box-framing aerogel configuration was developed \cite{Ref12}, which allows the use of the two-layer aerogel tile. With the box-framing aerogel tile, only the more robust 0.03 g cm$^{-3}$ layer is compressed on all four sides to fix the tile to the holder. In addition, the thickness of the 0.01 g cm$^{-3}$ surface layer was adjusted so as not to touch the grids during launch vibration. Moreover, the holder, notably the grid, was also redesigned to ensure that it does not touch the 0.01 g cm$^{-3}$ aerogel.

Furthermore, a prototype box-framing aerogel tile satisfied the safety requirements for payloads used in the ISS. The capture panels will be attached and detached to the ExHAM in the pressurized JEM by the ISS crews. During the flight, from rocket launch to retrieval into the pressurized module, the aerogel must not pose a danger to the ISS crew due to release from the holder. The box-framing aerogel passed the following qualification tests: (i) random vibration test for launch by the H-II Transfer Vehicle, Russian Progress/\\Soyuz, or SpaceX Dragon spacecraft, (ii) depressurization and repressurization test simulating the JEM airlock \cite{Ref12}, and (iii) thermal cycle test from $-$110$^\circ $C to 300$^\circ $C \cite{Ref24}. Extracted tiles from the aerogel products fabricated in this study passed the flight model acceptance tests. In addition, the feasibility of safe retrieval of the aerogel capture media was ensured. The outgassing of the aerogel, however, was beyond the scope of the safety requirements because of the small amount of aerogel tiles used per annual exposure (i.e., less than 1.9 L).

\subsection{Contamination control}
\label{sec:5.5}

The aerogel tiles for use in the Tanpopo experiment must be manufactured in contamination-controlled environments (for this study, the laboratory at Chiba University was used) because it is significantly difficult to distinguish the captured microbes/organic matters from ground-based ones (if they are contaminated in the aerogel production procedure). Aerogel tiles were fabricated basically in a clean booth (class 1,000), and sterile tools were used. An investigation of possible microbial contaminants in the manufacturing environment was previously performed \cite{Ref25}. In the previous study, deoxyribonucleic acid (DNA) contamination was not detected (i.e., below the detection limit) in an aerogel tile, indicating that it was produced in environments with adequate contamination control. In addition, fabrication techniques for reducing any amino acid contamination from human was adopted (personal communication from Mita H, Fukuoka Institute of Technology, Japan in 2011); e.g., the use of HPLC-grade ethanol and the supercritical CO$_2$ drying method instead of supercritical ethanol drying method. In the previous pilot production, we also practiced systematic mass production of aerogel tiles in contamination-controlled environments \cite{Ref12}. In the pilot production, the effective managements of resource and schedule were investigated. In the present study, these various techniques were applied during the production of the box-framing aerogels.

\subsection{Dust capture demonstration}
\label{sec:5.6}

The dust capture capability of the double-layer aerogels (and their surface layers) was investigated during a series of ground-based hypervelocity impact experiments using a two-stage light-gas gun at the ISAS Space Plasma Laboratory. In the first experiment, glass projectiles were used to represent perforative and stable dust grains. A prototype double-layer aerogel tile with densities of approximately 0.01 and 0.03 g cm$^{-3}$ successfully captured soda lime glass particles with a diameter of 30 $\mu $m that impacted the aerogel surface at an initial velocity of 6 km s$^{-1}$ \cite{Ref25}. The particles totally penetrated the 0.01 g cm$^{-3}$ surface layer and were stopped in the 0.03 g cm$^{-3}$ base layer, where the thickness of each layer was approximately 10 mm \cite{Ref25}. The demonstrated impact speed is possible in LEOs.

Next, powder particles separated from the Murchison meteorite were used in the impact experiment as brittle dust projectiles to simulate interplanetary dust particles. In this experiment, a single-layer 0.01 g cm$^{-3}$ aerogel block was impacted with 30--100 $\mu $m-diameter Murchison particles at a velocity of approximately 4 km s$^{-1}$ \cite{Ref26}. Micro-Fourier transform infrared and micro-Raman spectroscopy methods were established for analysis of the organic matter in meteoritic dust grains extracted from the impact tracks in the aerogel \cite{Ref26}. HPLC can also be used to analyze captured volatile dust grains without the need to extract them from the aerogel \cite{Ref9,Ref10}.

To develop a method for detecting microbes contained in the captured dust, particles made of smectite clay were also used as a model for terrestrial dust. Single- and double-layer aerogel blocks were impacted with approximately 60 $\mu $m-diameter smectite particles mixed with a specific microbe (\textit{Deinococcus radiodurans}) at approximately 4 km s$^{-1}$ \cite{Ref27}. Extracted aerogel segments containing the impact tracks and captured particles were subjected to a microbe detection procedure based on fluorescent microscopy using a DNA-specific fluorescent dye. This microbe identification method is one of the most promising technologies for the analysis of samples from the present space mission, and hence it was investigated in the ground-based experiment \cite{Ref27}; more detailed analysis will be performed elsewhere.

\section{Conclusion}
\label{sec:6}

Ultralow-density box-framing silica aerogel tiles were fabricated for use in the Tanpopo experiment performed on the ISS. The design of capture panels and the requirements for aerogel performance as both a cosmic dust capture medium and space flight hardware were clarified. The aerogel fabrication method, which was employed for the successful mass production of 60 aerogel tiles, was described in detail. In addition, the optimal density and dimensions for aerogel tiles were confirmed. Finally, the flight robustness, possible biochemical contamination, and dust capture performance of the aerogel tiles were demonstrated. Consequently, both the production method and box-framing aerogel tiles met the space mission requirements and are suitable for the Tanpopo experiment.

%
%


\begin{acknowledgements}
The authors are grateful to the members of the Tanpopo Collaboration and Prof. I. Adachi of the High Energy Accelerator Research Organization (KEK) in Japan for their assistance with the aerogel design, development, and fabrication. This study was supported in part by the Space Plasma Laboratory at ISAS, JAXA, and we would like to thank Dr. S. Hasegawa and the crew for their two-stage light-gas gun operation.
\end{acknowledgements}



\end{document}